\documentclass[10pt,a4paper,onecolumn]{article}
\usepackage{marginnote}
\usepackage{graphicx}
\usepackage{xcolor}
\usepackage{authblk,etoolbox}
\usepackage{titlesec}
\usepackage{calc}
\usepackage{tikz}
\usepackage{hyperref}
\hypersetup{colorlinks,breaklinks=true,
            urlcolor=[rgb]{0.0, 0.5, 1.0},
            linkcolor=[rgb]{0.0, 0.5, 1.0},
            citecolor = blue}
\usepackage{caption}
\usepackage{tcolorbox}
\usepackage{amssymb,amsmath}
\usepackage{ifxetex,ifluatex}
\usepackage{seqsplit}
\usepackage{xstring}
\usepackage{natbib}
\usepackage[T1]{fontenc}
\usepackage[utf8]{inputenc}
\usepackage{cleveref}
\usepackage{float}
\usepackage{orcidlink}
\usepackage{xspace}
\usepackage{listings}
\usepackage{comment}
\usepackage{markdown}
\usepackage{tikz}
\usetikzlibrary{shapes,arrows,shadows,fit}
\usetikzlibrary{positioning}
\usetikzlibrary{bayesnet}

\let\origfigure\figure
\let\endorigfigure\endfigure
\renewenvironment{figure}[1][2] {
    \expandafter\origfigure\expandafter[H]
} {
    \endorigfigure
}

\let\textttOrig=\texttt
\def\texttt#1{\expandafter\textttOrig{\seqsplit{#1}}}
\renewcommand{\seqinsert}{\ifmmode
  \allowbreak
  \else\penalty6000\hspace{0pt plus 0.02em}\fi}

\usepackage{abstract}

\makeatletter
\let\href@Orig=\href
\def\href@Urllike#1#2{\href@Orig{#1}{\begingroup
    \def\Url@String{#2}\Url@FormatString
    \endgroup}}
\def\href@Notdoi#1#2{\def\tempa{#1}\def\tempb{#2}%
  \ifx\tempa\tempb\relax\href@Urllike{#1}{#2}\else
  \href@Orig{#1}{#2}\fi}
\def\href#1#2{%
  \IfBeginWith{#1}{https://doi.org}%
  {\href@Urllike{#1}{#2}}{\href@Notdoi{#1}{#2}}}
\makeatother

\newlength{\cslhangindent}
\newlength{\csllabelwidth}
 {
  \setlength{\parindent}{0pt}
  \ifodd #1 \everypar{\setlength{\hangindent}{\cslhangindent}}\ignorespaces\fi
  \ifnum #2 > 0
  \setlength{\parskip}{#2\baselineskip}
  \fi
 }%
 {}
\usepackage{calc}

\usepackage[top=3.5cm, bottom=3cm, right=1.5cm, left=1.0cm,
            headheight=2.2cm, reversemp, includemp, marginparwidth=4.5cm]{geometry}

\titleformat{\section}
  {\normalfont\sffamily\Large\bfseries}
  {}{0pt}{}
\titleformat{\subsection}
  {\normalfont\sffamily\large\bfseries}
  {}{0pt}{}
\titleformat{\subsubsection}
  {\normalfont\sffamily\bfseries}
  {}{0pt}{}
\titleformat*{\paragraph}
  {\sffamily\normalsize}

\usepackage{fancyhdr}
\makeatletter
\let\ps@plain\ps@fancy
\fancyheadoffset[L]{4.5cm}
\fancyfootoffset[L]{4.5cm}

\definecolor{linky}{rgb}{0.0, 0.5, 1.0}

\newtcolorbox{repobox}
   {colback=red, colframe=red!75!black,
     boxrule=0.5pt, arc=2pt, left=6pt, right=6pt, top=3pt, bottom=3pt}

\newcommand{\ExternalLink}{%
   \tikz[x=1.2ex, y=1.2ex, baseline=-0.05ex]{%
       \begin{scope}[x=1ex, y=1ex]
           \clip (-0.1,-0.1)
               --++ (-0, 1.2)
               --++ (0.6, 0)
               --++ (0, -0.6)
               --++ (0.6, 0)
               --++ (0, -1);
           \path[draw,
               line width = 0.5,
               rounded corners=0.5]
               (0,0) rectangle (1,1);
       \end{scope}
       \path[draw, line width = 0.5] (0.5, 0.5)
           -- (1, 1);
       \path[draw, line width = 0.5] (0.6, 1)
           -- (1, 1) -- (1, 0.6);
       }
   }

\patchcmd{\@maketitle}{center}{flushleft}{}{}
\patchcmd{\@maketitle}{center}{flushleft}{}{}
\patchcmd{\@maketitle}{\LARGE}{\LARGE\sffamily}{}{}
\def\maketitle{{%
  \renewenvironment{tabular}[2][]
    {\begin{flushleft}}
    {\end{flushleft}}
  \AB@maketitle}}
\makeatletter
\renewcommand\AB@affilsepx{ \protect\Affilfont}
\renewcommand\AB@affilnote[1]{{\bfseries #1}\hspace{3pt}}
\renewcommand{\affil}[2][]%
   {\newaffiltrue\let\AB@blk@and\AB@pand
      \if\relax#1\relax\def\AB@note{\AB@thenote}\else\def\AB@note{#1}%
        \setcounter{Maxaffil}{0}\fi
        \begingroup
        \let\href=\href@Orig
        \let\texttt=\textttOrig
        \let\protect\@unexpandable@protect
        \def\thanks{\protect\thanks}\def\footnote{\protect\footnote}%
        \@temptokena=\expandafter{\AB@authors}%
        {\def\\{\protect\\\protect\Affilfont}\xdef\AB@temp{#2}}%
         \xdef\AB@authors{\the\@temptokena\AB@las\AB@au@str
         \protect\\[\affilsep]\protect\Affilfont\AB@temp}%
         \gdef\AB@las{}\gdef\AB@au@str{}%
        {\def\\{, \ignorespaces}\xdef\AB@temp{#2}}%
        \@temptokena=\expandafter{\AB@affillist}%
        \xdef\AB@affillist{\the\@temptokena \AB@affilsep
          \AB@affilnote{\AB@note}\protect\Affilfont\AB@temp}%
      \endgroup
       \let\AB@affilsep\AB@affilsepx
}
\makeatother

\renewcommand\Affilfont{\sffamily\small\mdseries}
\ifnum 0\ifxetex 1\fi\ifluatex 1\fi=0 
  \usepackage[T1]{fontenc}
  \usepackage[utf8]{inputenc}

\else 
  \ifxetex
    \usepackage{mathspec}
    \usepackage{fontspec}

  \else
    \usepackage{fontspec}
  \fi
  \defaultfontfeatures{Ligatures=TeX,Scale=MatchLowercase}

\fi
\IfFileExists{upquote.sty}{\usepackage{upquote}}{}
\IfFileExists{microtype.sty}{%
\usepackage{microtype}
\UseMicrotypeSet[protrusion]{basicmath} 
}{}

\let\addcontentslineOrig=\addcontentsline
\def\addcontentsline#1#2#3{\bgroup
  \let\texttt=\textttOrig\addcontentslineOrig{#1}{#2}{#3}\egroup}
\let\markbothOrig\markboth
\def\markboth#1#2{\bgroup
  \let\texttt=\textttOrig\markbothOrig{#1}{#2}\egroup}
\let\markrightOrig\markright
\def\markright#1{\bgroup
  \let\texttt=\textttOrig\markrightOrig{#1}\egroup}

\usepackage{graphicx,grffile}
\makeatletter
\def\maxwidth{\ifdim\Gin@nat@width>\linewidth\linewidth\else\Gin@nat@width\fi}
\def\maxheight{\ifdim\Gin@nat@height>\textheight\textheight\else\Gin@nat@height\fi}
\makeatother
\setkeys{Gin}{width=\maxwidth,height=\maxheight,keepaspectratio}
\IfFileExists{parskip.sty}{%
\usepackage{parskip}
}{
\setlength{\parindent}{0pt}
\setlength{\parskip}{6pt plus 2pt minus 1pt}
}
\ifx\paragraph\undefined\else
\let\oldparagraph\paragraph
\renewcommand{\paragraph}[1]{\oldparagraph{#1}\mbox{}}
\fi
\ifx\subparagraph\undefined\else
\let\oldsubparagraph\subparagraph
\renewcommand{\subparagraph}[1]{\oldsubparagraph{#1}\mbox{}}
\fi

\DeclareFixedFont{\ttb}{T1}{txtt}{bx}{n}{10} 
\DeclareFixedFont{\ttm}{T1}{txtt}{m}{n}{10}  

\usepackage{color}
\definecolor{deepblue}{rgb}{0,0,0.5}
\definecolor{deepred}{rgb}{0.6,0,0}
\definecolor{deepgreen}{rgb}{0,0.5,0}

\usepackage{listings}

\newcommand\pythonstyle{\lstset{
language=Python,
basicstyle=\scriptsize,
morekeywords={self},              
keywordstyle=\scriptsize\color{deepblue},
emph={MyClass,__init__},          
emphstyle=\ttb\color{deepred},    
stringstyle=\color{deepgreen},
    numbers=left,
frame=tb,                         
showstringspaces=false
}}

\lstnewenvironment{python}[1][]
{
\pythonstyle
\lstset{#1}
}
{}

\newcommand\pythoninline[1]{{\pythonstyle\lstinline!#1!}}

\newcommand{\pkgfont}{\texttt} 
\newcommand{\pkg}[2][]{%
  \if\relax\detokenize{#1}\relax
    \pkgfont{#2}%
  \else
    \href{#1}{\pkgfont{#2}}%
  \fi
}

\newcommand{\smokescreen}{\pkg{Smokescreen}}
\newcommand{\firecrown}{\pkg{Firecrown}}
\newcommand{\cosmosis}{\pkg{CosmoSIS}}
\newcommand{\ccl}{\pkg{pyCCL}}
\newcommand{\sacc}{\pkg{SACC}}

\usepackage{listings}
\usepackage{xcolor}

\lstdefinelanguage{YAML}{
  keywords={true, false, null},
  keywordstyle=\color{blue}\bfseries,
  comment=[l]{\#},
  commentstyle=\color{gray}\itshape,
  stringstyle=\color{teal},
  morestring=[b]",
  morestring=[b]',
}

\newcommand{\threextwo}{3$\times$2pt}

\title{\texttt{Smokescreen}: A Python package for data vector blinding and encryption in cosmological analyses}

\date{\vspace{-7ex}}

\begin{document}
\author[1,2]{Arthur Loureiro\thanks{Corresponding author: arthur.loureiro@fysik.su.se}\orcidlink{0000-0002-4371-0876}}
\author[3]{Jessica Muir\orcidlink{0000-0002-7579-770X}}
\author[4]{Jonathan Blazek\orcidlink{0000-0002-4687-4657}}
\author[5,6]{Nora Elisa Chisari\orcidlink{0000-0003-4221-6718}}
\author[7,8]{Pedro H. Costa Ribeiro\orcidlink{0009-0007-9603-8335}}
\author[9]{Christos Georgiou\orcidlink{0000-0001-9707-0109}}
\author[10]{C. Danielle Leonard\orcidlink{0000-0002-7810-6134}}
\author[11,7]{Bruno Moraes\orcidlink{0000-0002-5898-0975}}
\author[12]{Marc Paterno\orcidlink{0000-0003-0808-8388}}
\author[13]{Nikolina \v{S}ar\v{c}evi\'c\orcidlink{0000-0001-7301-6415}}
\author[14]{Tilman Tröster\orcidlink{0000-0003-3520-2406}}
\author[8]{Sandro D. P. Vitenti\orcidlink{0000-0002-4587-7178}}
\author[ ]{the LSST Dark Energy Science Collaboration}

\affil[1]{Oskar Klein Centre for Cosmoparticle Physics, Department of Physics, Stockholm University, Stockholm, SE-106 91, Sweden}
\affil[2]{ Astrophysics Group, Blackett Laboratory, Imperial College London, London SW7 2AZ, UK}
\affil[3]{University of Cincinnati, Cincinnati, Ohio 45221, USA}
\affil[4]{Department of Physics, Northeastern University, Boston, MA 02115, USA}
\affil[5]{Institute for Theoretical Physics, Utrecht University, Princetonplein 5, 3584 CC, Utrecht, the Netherlands}
\affil[6]{Leiden Observatory, Leiden University, Niels Bohrweg 2, 2333 CA, Leiden, the Netherlands}
\affil[7]{Instituto de Física, Universidade Federal do Rio de Janeiro, Cidade Universitária, Rio de Janeiro, 21941-909, Brazil}
\affil[8]{Departamento de Física, Universidade Estadual de Londrina, Rod. Celso Garcia Cid, Km 380, 86057 970, Londrina, Paran\'a, Brazil}
\affil[9]{Institut de Física d’Altes Energies (IFAE), The Barcelona Institute of Science and Technology, Campus UAB, 08193 Bellaterra (Barcelona), Spain}
\affil[10]{School of Mathematics, Statistics and Physics, Newcastle University, Newcastle upon Tyne, NE1 7RU, United Kingdom}
\affil[11]{CBPF - Centro Brasileiro de Pesquisas Físicas, 22290-180, Rio de Janeiro, RJ, Brazil}
\affil[12]{Fermi National Accelerator Laboratory, Batavia, IL 60510-0500, U.S.A.}
\affil[13]{Department of Physics, Duke University, Science Dr, Durham, NC 27710, USA}
\affil[14]{Institute for Particle Physics and Astrophysics, ETH Zurich, 8093 Zurich, Switzerland}

\maketitle

\marginpar{
  \begin{flushleft}
  \sffamily\small

  {\bfseries DOI:} \href{https://doi.org/DOI\_TBD}{\color{linky}{DOI TBD}}

  \vspace{2mm}

  {\bfseries Software}
  \begin{itemize}
    \setlength\itemsep{0em}
    \item \href{N/A}{\color{linky}{Review}} \ExternalLink
    \item \href{https://github.com/LSSTDESC/Smokescreen/tree/main}{\color{linky}{Repository}} \ExternalLink
  \end{itemize}

  \vspace{2mm}
  \par\noindent\hrulefill\par
  \vspace{2mm}

  {\bfseries Editor:} \href{https://example.com}{Pending editor} \ExternalLink \\
  \vspace{1mm}
  {\bfseries Reviewers:}
  \begin{itemize}
    \setlength\itemsep{0em}
    \item \href{https://github.com/pending}{@pending}
  \end{itemize}

  \vspace{2mm}
  {\bfseries Submitted:} N/A \\
  {\bfseries Published:} N/A

  \vspace{2mm}
  {\bfseries License}\\
  Authors retain copyright and release the work under CC BY 4.0
  (\href{http://creativecommons.org/licenses/by/4.0/}{\color{linky}{link}}).
  \end{flushleft}
}

\vspace{.2cm}

\vspace{.5cm}

\section{Summary}
\noindent
\smokescreen\ is an open-source Python library for data-vector concealment (blinding) in cosmological analyses. 
Data-vector blinding works by applying cosmology-dependent shifts to the observed data vector, moving it away from the true cosmological signal without affecting its statistical properties, so that analysts cannot infer the true result until the analysis is frozen and the blinding is lifted. 
The package computes these shifts using \firecrown{}\footnote{\url{https://firecrown.readthedocs.io/en/latest/}} likelihoods applied to data vectors stored in the \sacc{}  format\footnote{\url{https://sacc.readthedocs.io/en/latest/intro.html}}, ensuring that the theoretical model used for blinding is identical to that used for inference whilst remaining agnostic to the specific observable being blinded. 
To prevent accidental unblinding, the original \sacc{} file, containing the true cosmology, is encrypted. 
Although developed for the Vera C. Rubin Observatory Legacy Survey of Space and Time (LSST), \smokescreen{} is applicable to any experiment using \firecrown{} likelihoods and the \sacc{} data format.

\section{Statement of Need}
\label{sec:state_of_need}

Modern cosmological analyses require robust data concealment (blinding) strategies to prevent experimenter bias.
Cosmological experiments are particularly challenging to blind due to a fundamental limitation: we can only observe one Universe, making it extremely hard to design analyses in a double-blind or even single-blind manner.
Additionally, cosmological signals are not localized events but statistical patterns distributed across large data sets such as correlation functions or power spectra, leaving no clear "signal region" that can simply be hidden.
Without a principled approach, analysts risk unconsciously tuning their pipelines towards an expected result or towards the results of other experiments, compromising the scientific integrity of the measurement.

\citet{Muir2020} formalised this discussion by defining three criteria for a reliable data concealment strategy: \textbf{(I)} the concealment must hide the true results while producing a physically reasonable posterior that preserves known parameter degeneracies; \textbf{(II)} it must retain the ability to perform validation tests, including goodness-of-fit and systematic null tests; and \textbf{(III)} it must be straightforward to implement and unblind, while still preventing accidental unblinding.
\citet{Muir2020} further showed that applying cosmology-dependent shifts in data vector space satisfies all three requirements.

\smokescreen{} follows this approach: given a reference cosmology and a randomly-drawn blinded cosmology, it computes the difference between the corresponding theory predictions and applies the shift to the observed data vector.
With LSST Y1 data forthcoming and DESC analyses \citep{2012-LSST_DESC} requiring a collaboration-wide blinding strategy, no existing tool in the DESC software ecosystem (or outside it) provided a standardised implementation of this method.
\smokescreen\ was developed to fill this gap.

\section{State of the Field}
\label{sec:state_of_field}
Blinding in cosmological surveys can be implemented at several stages of the analysis pipeline.
Catalogue-level methods \citep[e.g.,][]{2016MNRAS.460.2245J, 2020JCAP...09..052B} provide robust concealment but are difficult to implement, particularly for multi-probe experiments.
Posterior-level methods \citep[e.g.,][]{2006ApJ...644....1C} are simple but leave the data vectors unblinded throughout the analysis.
Covariance-level blinding \citep{Sellentin2020} is mathematically elegant but harder to apply in practice, as covariance matrices in multi-probe analyses are constructed in a variety of ways --- analytically, from simulations, or via resampling methods, for example. 
Data vector blinding \citep{Muir2020} provides a practical compromise, offering stronger guarantees than posterior-level methods while remaining easier to implement than catalogue- or covariance-level methods.

The Dark Energy Survey \citep{2026-DES}, Kilo Degree Survey \citep{2025-KiDS_Legacy}, and Hyper Suprime-Cam \citep{2023-HSC} have used combinations of these strategies in their analyses, but these implementations were largely \textit{ad hoc} and not easily reusable.
\smokescreen\ provides a dedicated open-source library for data vector blinding designed to integrate with collaboration-wide analysis pipelines.
By making the concealment procedure transparent and reproducible, it also allows the wider community to inspect and verify how blinding shifts are computed prior to and post unblinding.

\section{Software Design}
\label{sec:software}
\smokescreen{} is structured around a single primary abstraction, the \texttt{ConcealDataVector} class, with thin command-line and encryption wrappers layered on top. 
The design follows a separation-of-concerns principle keeping four responsibilities independent: \textbf{(I)} cosmological parameter perturbation (\texttt{param\_shifts.py}), \textbf{(II)} theory vector computation via an external likelihood framework (\texttt{datavector.py}), \textbf{(III)} data security (\texttt{encryption.py}), and \textbf{(IV)} user-facing orchestration (\texttt{\_\_main\_\_.py}). 
This structure allows any \firecrown{}\footnote{\firecrown{} is DESC's likelihood library, see \url{https://github.com/LSSTDESC/firecrown} for more information.}-compatible likelihood to be used, guaranteeing that the theoretical modelling used for concealment is identical to that used for inference, preserving Criterion~II of \citet{Muir2020}. \smokescreen{} can perform shifts in any of the base cosmological parameters present in \ccl{} \citep{2019-Chisari-CCL}.

Smokescreen's test suite is implemented using \pkg{pytest} and currently achieves 94\% code coverage as tracked by \pkg{Codecov}. 
Rather than merely checking array shapes or return types, tests verify physical correctness: for instance, that additive and multiplicative blinding factors are computed exactly as prescribed by Eqs.~\ref{eq:theory_vecs}--\ref{eq:factors}, that \sacc{} consistency checks raise errors on data vector or covariance mismatches to floating-point tolerance, that invalid parameter keys or shift distributions are caught at instantiation, and that the full concealment pipeline, from shift sampling through to the saved output file and its metadata, produces the expected result end-to-end.

\subsection{Core Blinding Algorithm}

\smokescreen{} implements the data vector blinding methodology of \citet{Muir2020} in five steps.
First, for each cosmological parameter $\theta_i$ the user specifies the 
parameters of a shift distribution (deterministic, flat uniform, or Gaussian; see \textit{Parameter Shift Strategies} below) from which $\Delta\theta_i$ is drawn. 
Second, a concealed cosmology is constructed:
\begin{equation}
    \boldsymbol{\theta}_{\rm blind} = \boldsymbol{\theta}_{\rm ref} + 
    \boldsymbol{\Delta\theta}.
    \label{eq:blind_cosmo}
\end{equation}
where $\theta_{\rm ref}$ is a reference cosmology defined by the user. Third, the  theory vector, $T(\boldsymbol{\theta},\boldsymbol{s})$, is evaluated by the \firecrown{} likelihood object at both cosmologies with nuisance parameters 
$\boldsymbol{s}_{\rm ref}$ fixed at their reference values:
\begin{equation}
    \mathbf{d}_{\rm ref} = T\!\left(\boldsymbol{\theta}_{\rm ref},\,
    \boldsymbol{s}_{\rm ref}\right), \qquad
    \mathbf{d}_{\rm blind} = T\!\left(\boldsymbol{\theta}_{\rm blind},\,
    \boldsymbol{s}_{\rm ref}\right).
    \label{eq:theory_vecs}
\end{equation}
Fourth, a blinding factor is computed in either additive or multiplicative form:
\begin{equation}
    f^{\rm add} = \mathbf{d}_{\rm blind} - \mathbf{d}_{\rm ref}, \qquad
    f^{\rm mult} = \mathbf{d}_{\rm blind} \,/\, \mathbf{d}_{\rm ref},
    \label{eq:factors}
\end{equation}
applied according to the observable. Fifth, the factor is applied to the measured data vector:
\begin{equation}
    \hat{\mathbf{d}} = \mathbf{d}_{\rm obs} + f^{\rm add} \quad \text{or} \quad
    \hat{\mathbf{d}} = \mathbf{d}_{\rm obs} \cdot f^{\rm mult}.
    \label{eq:blind_datavec}
\end{equation}
Fixing $\boldsymbol{s}_{\rm ref}$ in both evaluations ensures the blinding factor carries only the cosmological shift, leaving the covariance matrix unchanged and allowing standard inference pipelines to run on $\hat{\mathbf{d}}$ without 
modification.

\subsection{Software Architecture}

\begin{figure}
    \centering
    \includegraphics[width=\linewidth]{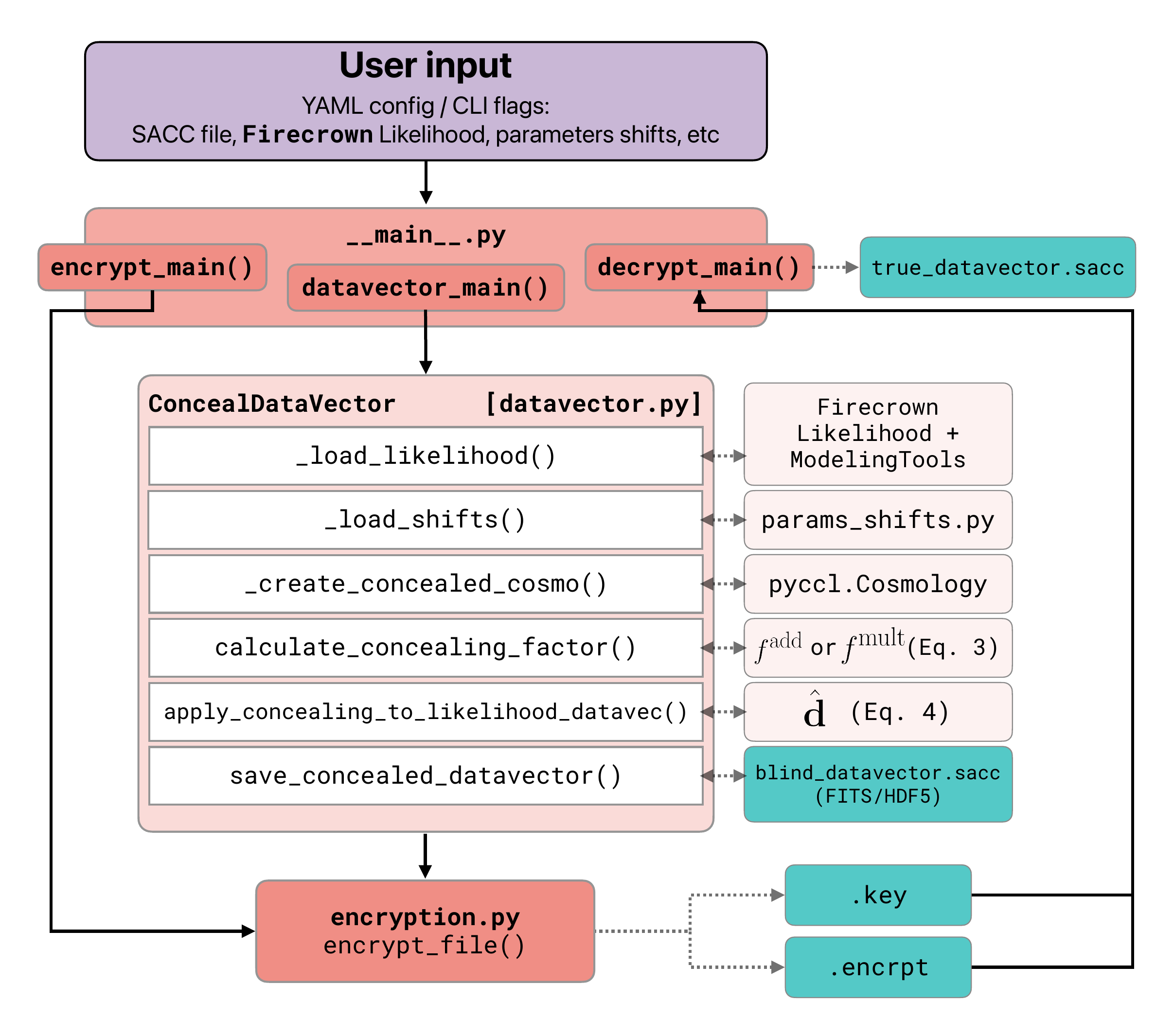}
    \caption{\smokescreen{} data-flow architecture. Solid arrows show the main execution sequence; dashed arrows show outputs produced by each step; green boxes show outputs. \texttt{datavector\_main()} orchestrates \texttt{ConcealDataVector}, which triggers encryption; \texttt{encrypt\_main()} provides a standalone encryption path; \texttt{decrypt\_main()} restores the original \sacc{} file given the \texttt{.encrpt} and \texttt{.key} files.
    }
    \label{fig:architecture}
\end{figure}

Figure~\ref{fig:architecture} shows the workflow as a sequential data-flow diagram.
The \texttt{ConcealDataVector} class orchestrates the process: it evaluates the theory vectors (Eqs.~\ref{eq:theory_vecs}--\ref{eq:factors}), applies the blinding factor (Eq.~\ref{eq:blind_datavec}), and writes the output \sacc{} file\footnote{The \sacc{} file format stores correlations, covariances, and associated metadata for cosmological analyses}.

Two validation guards prevent common configuration errors.
\texttt{\_verify\_sacc\_consistency()} checks the return \sacc{} from the likelihood matches the user's input to floating-point tolerance ($10^{-10}$), while \texttt{\_check\_amplitude\_parameter()} ensures consistent specification of the amplitude parameters $A_s$/$\sigma_8$ across the reference cosmology, \firecrown's \texttt{CCLFactory}\footnote{\firecrown{} interfaces with the DESC Core Cosmology Library, \ccl{} \citep{2019-Chisari-CCL}.}, and the shift dictionary.

Supporting modules handle parameter shift generation (\texttt{param\_shifts.py}, implemented as a stateless module for independent unit testing and modularity), symmetric encryption via \texttt{Fernet}\footnote{\url{https://github.com/pyca/cryptography}} (\texttt{AES-128-CBC + HMAC-SHA256} with a fresh key per blinding operation) (\texttt{encryption.py}, also exposed through command line interface (CLI) subcommands), and \sacc{} format auto-detection with cosmology construction (\texttt{utils.py}).

\subsection{\firecrown{} Likelihood Integration}

\smokescreen{} requires only a Python module exporting a \texttt{build\_likelihood(build\_parameters)} factory returning a \texttt{(Likelihood, ModelingTools)} tuple conforming to the \firecrown{} interface --- the same interface used by the \firecrown{} with the \texttt{CosmoSIS}/\texttt{Cobaya}/\texttt{NumCosmos} \citep[respectively]{2015-Cosmosis, 2021-Cobaya, Vitenti2012c} connectors. 
Therefore, existing likelihood modules do not require modification.

\subsection{Parameter Shift Strategies} \label{sec:param_shifts}

Three shift distributions are implemented: deterministic ($\theta_i^{\rm blind} = \rm{const}$), uniform ($\theta_i^{\rm blind} \sim \mathcal{U}(a,b)$), and Gaussian ($\theta_i^{\rm blind} \sim \mathcal{N}(\mu,\sigma)$).
Reproducibility is ensured through integer or string seeds, with strings mapped to \texttt{NumPy}-compatible integers via \texttt{MD5} hashing.
Unknown parameter keys raise a \texttt{ValueError} during validation, preventing silent misapplication of shifts.

\subsection{Security and Data Integrity}

After the concealed \sacc\ file is written, the original is encrypted with \texttt{Fernet} and deleted by default.
The \texttt{.key} file is stored separately from the \texttt{.encrpt} data file, allowing a blinding committee to distribute them through independent channels.
The blinded \sacc\ embeds audit metadata (seed, timestamp, username) sufficient to reconstruct the blinding configuration without external logs.

\subsection{CLI and Configuration}
\smokescreen{} exposes three subcommands --- \texttt{datavector}, \texttt{encrypt}, and \texttt{decrypt} --- implemented with \texttt{jsonargparse}\footnote{\url{https://github.com/omni-us/jsonargparse/}}.
Arguments may be provided as command-line interface (CLI) flags or through a \texttt{YAML} configuration file (\texttt{--config\,\, config.yaml}); \texttt{--print\_config} displays the current argument schema.
The \texttt{reference\_cosmology} argument accepts a partial parameter dictionary, completed using \texttt{pyccl.CosmologyVanillaLCDM} \citep{2019-Chisari-CCL} defaults.
Input/output format (\texttt{FITS} or \texttt{HDF5}) is auto-detected and preserved.

\section{Application: LSST Y1 \threextwo\ Concealment}
\label{sec:application}

We demonstrate \smokescreen{} on a simulated LSST Y1 \threextwo\ data vector comprising cosmic shear $C_\ell^{\gamma\gamma}$, galaxy--galaxy lensing $C_\ell^{g\gamma}$, and galaxy clustering $C_\ell^{gg}$  \citep{2023OJAp....6E..13P} across five tomographic redshift bins, yielding 415 data points sampled at 20 log-spaced multipoles $\ell \in [20, 2000]$ with the samples containing galaxy clustering restricted to scales $k\leq 0.1$ Mpc$^{-1}$.

\subsubsection{\smokescreen{} Configuration}

Two blinded realisations (``Blind A'' and ``Blind B'') were produced with deterministic $A_s$ and $w$ shifts from the fiducial value $A_{s}^{\rm fid} = 1.9019 \times 10^{-9}$, and $w^{\rm fid}=-1.0$.

Blind A settings are:
\begin{lstlisting}[language=YAML]
path_to_sacc: "./sacc_forecasting_y1_3x2pt.sacc"
likelihood_path: "./3x2pt_likelihood.py"
output_suffix: "blind_A"
shifts_dict:
    A_s: 2.00e-09  # +5.2% shift from fiducial
    w: -1.1  # +10.0% shift from fiducial
shift_distribution: "flat"
systematics:
    lens0_bias: 1.2497
    lens1_bias: 1.3809
    lens2_bias: 1.5231
    lens3_bias: 1.6716
    lens4_bias: 1.8245
    ia_bias: 1.0
    alphaz: 0.0
    z_piv: 0.62
reference_cosmology:
    A_s: 1.9019e-09
    Omega_c: 0.2906
keep_original_sacc: true
\end{lstlisting}

While Blind B uses:
\begin{lstlisting}[language=YAML]
output_suffix: "blind_B"
shifts_dict:
    A_s: 1.80e-09  # -5.2% shift from fiducial
    w: -0.9  # -10.0% shift from fiducial
\end{lstlisting}
and is otherwise identical. 

\begin{figure}
    \centering
    \includegraphics[width=\linewidth]{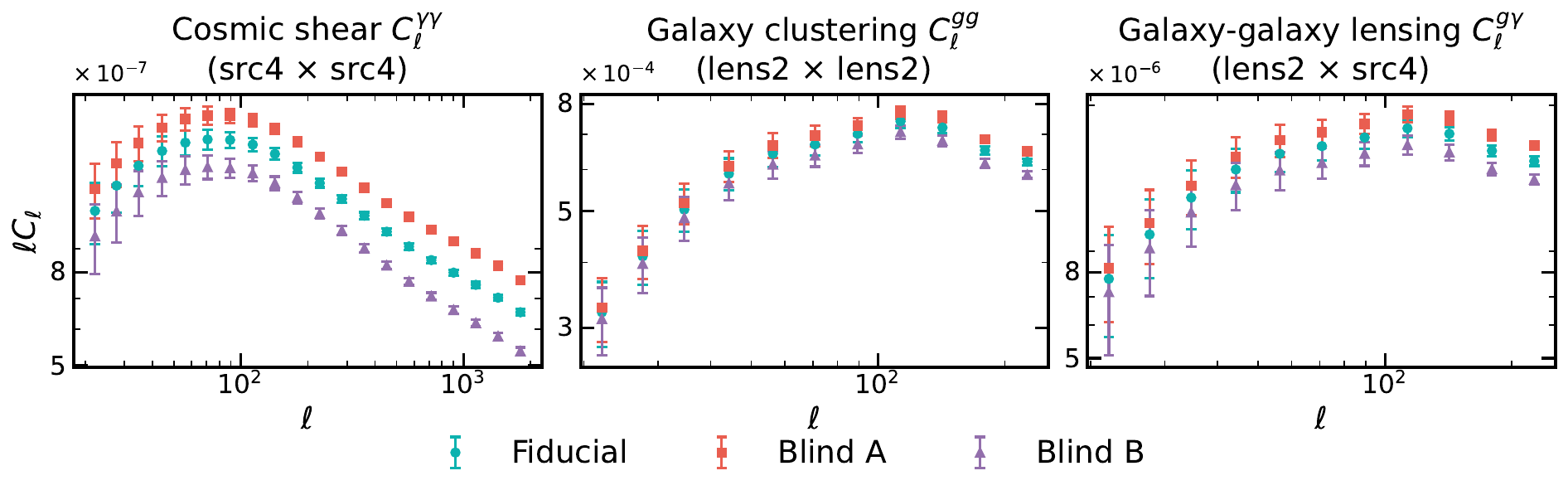}
    \caption{Example of multi-probe concealment for a simulated LSST Y1 \threextwo\ data vector: (left) cosmic shear, (middle) galaxy clustering, and (right) galaxy--galaxy lensing.
    Green circles show the original data vector ($A_s=1.9\times10^{-9}$, $w=-1.0$).
    Red squares (purple triangles) show \textit{Blind A} (\textit{Blind B}), produced by
    deterministic shifts to $A_s=2.0\times10^{-9}$, $w=-1.1$ ($A_s=1.8\times10^{-9}$, $w=-0.9$).
    }
    \label{fig:data-vectors}
\end{figure}

In a non-determinist shift case, the user would specify a list with two values, e.g. the limits of a uniform distribution and a seed:
\begin{lstlisting}[language=YAML]
shifts_dict:
    A_s: [1.70e-09, 2.20e-09]
    w: [-1.3, -0.8]
shift_distribution: "flat"
seed: 2112
\end{lstlisting}

Finally, the concealment of the data vector is invoked as:
\begin{verbatim}
smokescreen datavector --config conceal_lsst_y1_3x2pt_blind_[A/B].yaml
\end{verbatim}

producing \texttt{sacc\_forecasting\_y1\_3x2pt\_blind\_[A/B].sacc} and an
encrypted copy of the original. The resulting concealed data vectors are shown in Figure~\ref{fig:data-vectors}.

\subsection{Cosmological Validation}
Following \citet{Muir2020}, valid concealment shifts should move the posterior best-fit sufficiently to blind the posterior while leaving the goodness-of-fit unchanged (Criterion~II). 
We verify this with a lightweight inference over three cosmological parameters: $A_s$, $w$, and $\Omega_{\rm cdm}$ only\footnote{Note that only $A_s$ and $w$ are concealed in this test, while $\Omega_{\rm cdm}$ is kept the same.}, using \cosmosis\ \citep{2015-Cosmosis} and the same \firecrown\ likelihood supplied to \smokescreen.
Deterministic shifts allow direct comparison with the known input cosmology.
A full validation on realistic simulated analyses, including the complete parameter space and systematics, will be presented in a forthcoming LSST DESC study.

\begin{figure}
    \centering
    \includegraphics[width=0.9\linewidth]{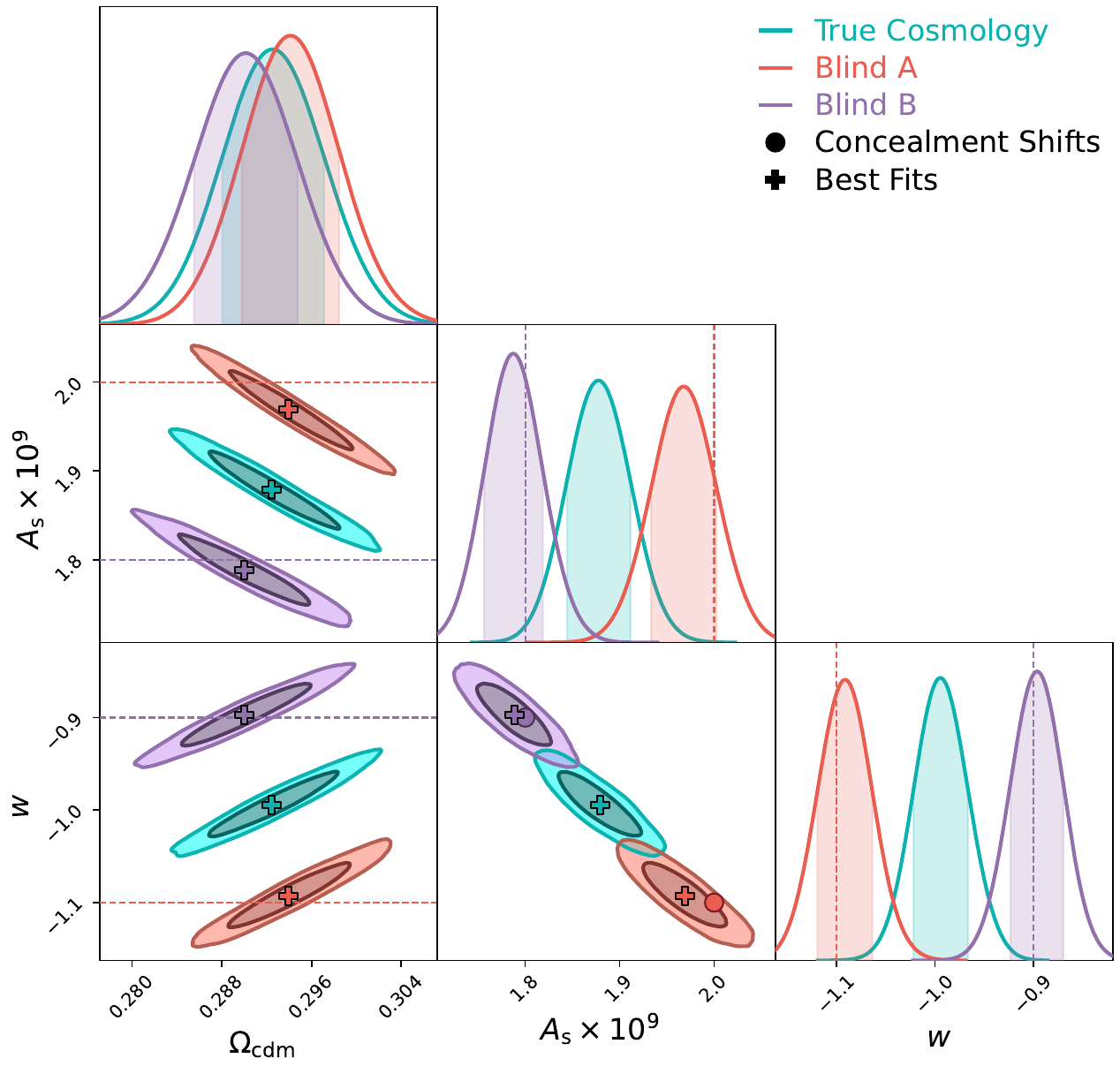}
    \caption{Marginalised 1D and 2D posterior distributions from cosmological inference on the data vectors shown in Figure~\ref{fig:data-vectors}, using the same \firecrown{} likelihood provided to \smokescreen{}. 
    The true cosmology (cyan), Blind~A (red), and Blind~B (purple) posteriors shift consistently towards their respective input concealing cosmologies, marked by filled circles and dashed lines.
    Cross markers indicate the posterior best-fits reported in Table~\ref{tab:evidence}.
    Note that $\Omega_{\rm cdm}$ was not shifted by  \smokescreen{}, yet its posterior shifts a tiny amount as a consequence of the known  $A_s$--$\Omega_{\rm cdm}$--$w$ degeneracy, demonstrating that Criterion~I is satisfied.}
    \label{fig:triangle}
\end{figure}

Figure~\ref{fig:triangle} confirms the expected behaviour: posteriors shift towards the blinded cosmology, whilst the goodness-of-fit remains unchanged ($\chi^2_{\rm red} = 0$ for the noiseless fiducial vector). 
The Bayesian log-evidence $\log Z$, computed via \texttt{Nautilus} \citep{nautilus}, is likewise unaffected: $|\Delta\log Z| \leq 1$ for both blinded runs (Blind~A: $-0.97$, Blind~B: $-0.47$), which is inconclusive on the Jeffreys scale and confirms the blinded analyses are statistically indistinguishable from the unblinded one. The results are summarised in Table~\ref{tab:evidence}.

\begin{table}[h]
    \centering
    \caption{Best fit values, goodness-of-fit, and log-evidence for all three cases in Figure~\ref{fig:triangle}.}
    \label{tab:evidence}
    \begin{tabular}{l|c|c|c|c|c|}
                       &  $A_s\times10^{-9}$    &  $w$  &  $\Omega_{\rm cdm}$ & $\chi_{\rm red}^2$ & $\log Z$ \\
        \hline
        True Cosmology & $1.88$& $-0.99$         & $0.292$ & $0.0039$ & $-16.67 \pm 0.03$ \\
        Blind A        & $1.97$& $-1.09$         & $0.294$ & $0.0084$ & $-17.64 \pm 0.03$ \\
        Blind B        & $1.79$& $-0.90$         & $0.290$ & $0.0069$ & $-17.14 \pm 0.03$ \\
    \end{tabular}
\end{table}

\section{Research Impact Statement}
\label{sec:research_impact}

\smokescreen\ is designed as a standard blinding infrastructure for LSST DESC cosmological analyses ahead of LSST’s decade-long survey.
It eliminates inconsistencies between blinding and inference models that can arise from independent \textit{ad hoc} implementations across analysis teams.
Adoption is already extending beyond LSST DESC: the KiDS collaboration is using \smokescreen\ to blind their upcoming $6\times2$pt legacy analysis, demonstrating its applicability to any experiment using \firecrown\ likelihoods and the \sacc\ format. 
More broadly, we hope \smokescreen\ will help drive the community towards more consistent, open, and verifiable blinding infrastructures ahead of the Stage-IV survey era.

\section{Availability}
\noindent
\textbf{Source:} \href{https://github.com/LSSTDESC/Smokescreen/tree/main}{github.com/LSSTDESC/Smokescreen} \\
\textbf{License:} BSD 3-Clause License. \\
\textbf{Install (conda):} \texttt{conda\,\, install\,\, -c\,\, conda-forge\,\, lsstdesc-smokescreen} \\
\textbf{Install (PyPI):} \texttt{pip\,\, install\,\, smokescreen} \\
\textbf{Documentation:} \href{ https://lsstdesc.org/Smokescreen/}{lsstdesc.org/Smokescreen/}\\
\textbf{Examples:}
\href{https://github.com/LSSTDESC/Smokescreen/tree/main/examples}{github.com/LSSTDESC/Smokescreen/tree/main/examples}\\
\textbf{Notebooks:}
\href{https://github.com/LSSTDESC/Smokescreen/tree/main/notebooks}{github.com/LSSTDESC/Smokescreen/tree/main/notebooks}

\section{AI usage disclosure}
Generative AI tools, including GitHub Copilot and Claude Code, played a minor supporting role in the development of \smokescreen, primarily in the generation of unit tests and in resolving small, localised bugs. 
All AI-assisted code was reviewed and edited by the authors before being incorporated into the codebase. 
The core design of the library, its scientific grounding, and all key implementation decisions were made entirely by the authors. 
The writing in this paper is our own; AI assistance was limited to occasional grammar corrections and light rephrasing to improve fluency, as several of the authors are non-native English speakers.

\section*{Acknowledgments}
 This paper has undergone internal review in the LSST Dark Energy Science Collaboration. We thank Joe Zuntz and Jayme Ruiz-Zapatero for acting as DESC internal reviewers for this work.
 AL acknowledges support from the Swedish National Space Agency (Rymdstyrelsen) under Career Grant Project Dnr 2024-00171 and from the research project grant `Understanding the Dynamic Universe' funded by the Knut and Alice Wallenberg Foundation under Dnr KAW 2018.0067. \smokescreen{} was developed as a part of the Stockholm University in-kind software contributions to the Vera Rubin Observatory's Legacy Survey of Space and Time. CG is funded by the MICINN project PID2022-141079NB-C32. IFAE is partially funded by the CERCA program of the Generalitat de Catalunya. N\v{S} is supported in part by the OpenUniverse effort, which is funded by NASA under JPL Contract Task 70-711320, ‘Maximizing Science Exploitation of Simulated Cosmological Survey Data Across Surveys.’ TT acknowledges funding from the Swiss National Science Foundation under the Ambizione project PZ00P2\_193352

\section*{Software Acknowledgments}
This project is largely possible thanks to several key software tools and packages. 
We gratefully acknowledge the Python programming language, without which this work would not have been possible.\footnote{\url{https://www.python.org}}
Our development and analysis efforts relied heavily on essential Python libraries, including: \pkg{NumPy} \citep{numpy}, \pkg{SciPy} \citep{scipy}, \pkg{Matplotlib} \citep{matplotlib}, \pkg{ChainConsumer} \citep{Hinton2016}, \pkg{AstroPy} \citep{astropy:2013, astropy:2018, astropy:2022}, HDF5 \citep{andrew_collette_2022_6575970} 
and cosmology packages like \ccl{} \citep{2019-Chisari-CCL}, \pkg{CosmoSIS} \citep{2015-Cosmosis},
\pkg{CAMB} \citep{2000ApJ...538..473L,2012JCAP...04..027H}, \pkg{HMCode2020} \citep{2021MNRAS.502.1401M}, and \pkg{NaMaster} \citep{2019MNRAS.484.4127A}.

We also acknowledge the use of \pkg{Jupyter\,\, Notebooks} \citep{jupyter} as an interactive computing environment for prototyping and documenting analyses.

\section*{Author Contributions}
We outline the different contributions below using keywords based on the Contribution Roles Taxonomy (CRediT; \citealp{brand15}). 
\textbf{AL (core developer \& maintainer of \smokescreen{}):} 
Conceptualization, Methodology, Software, Validation, Formal analysis, 
Investigation, Resources, Writing - Original Draft, Visualization, Project administration;
\textbf{JM:} Conceptualization, Methodology, Project administration;
\textbf{JB:} Conceptualization, Project administration;
\textbf{NEC:} LSST DESC Builder, contributor to {\tt pyCCL};
\textbf{PHCR:} Data Curation, Software*;
\textbf{CG:} Software; 
\textbf{CDL:} LSST DESC Builder, contributor to {\tt pyCCL};
\textbf{BM:} Software*;
\textbf{MP:} Software;
\textbf{N\v{S}:} Writing - Review \& Editing, Visualization;
\textbf{TT:} LSST DESC Builder, contributor to {\tt pyCCL} \& {\tt Firecrown}.
\textbf{SPDV:} Software*;

* not direct \smokescreen{} development, but software used in the work presented here.

\bibliographystyle{mnras}
\bibliography{main_joss}

\end{document}